\documentclass[pre,twocolumn,twoside,byrevtex,superscriptaddress,floatfix,showpacs,footinbib]{revtex4-1}

\usepackage{currfile}
\usepackage{color}

\usepackage{graphicx,epsfig,verbatim,enumerate}
\usepackage{amssymb,amsmath}
\usepackage{ifthen}

\usepackage{longtable}

\usepackage{mathtools}

\newboolean{twocolswitch}

\newcommand{\avg}[1]{\left\langle#1\right\rangle}

\newcommand{\sindex}[1]{}
\newcommand{\nindex}[1]{}

\newcommand{\www}[1]{\url{#1}}

\newcommand{\Req}[1]{Eq.~(\ref{#1})}

\definecolor{linkgrey}{RGB}{100, 100, 100}

\usepackage{lettrine}

\newcommand{\dee}[1]{\mbox{d}#1}

\newcommand{\Prob}[1]{{\rm P}\left(#1\right)}

\newcommand{\prob}[1]{{\rm P}(#1)}
\newcommand{\probccdf}[1]{{\rm P}_{\ge}(#1)}

\newcommand{\zipfexponent}{\alpha}

\newcommand{\zipfrank}{r}
\newcommand{\groupnum}{n}

\newcommand{\innovationrate}{\rho}
\newcommand{\innovationratet}{\rho_{t}}

\newcommand{\rgrSize}{S}
\newcommand{\Ngroups}{N}

\newcommand{\rgrSizeinit}{\rgrSize^{\textnormal{\textnormal{init}}}}

\newcommand{\grouptimestart}[1]{t_{#1}^{\textnormal{init}}}

\newcommand{\groupsizet}[2]{\rgrSize_{#1,#2}}

\newcommand{\groupnumt}[1]{\Ngroups^{\textnormal{(g)}}_{#1}}

\newcommand{\rgrP}{P}
\newcommand{\groupprobt}[2]{\rgrP_{#1,#2}}

\newcommand{\Beta}{B}

\newcommand{\groupnuminit}{\groupnum^{\textnormal{init}}}

\setboolean{twocolswitch}{true}

\begin{document}

\title{\protect
Simon's fundamental rich-get-richer model entails a dominant first-mover advantage
}

\author{
  \firstname{Peter Sheridan}
  \surname{Dodds}
}

\email{peter.dodds@uvm.edu}

\affiliation{
  Vermont Complex Systems Center,
  Computational Story Lab,
  the Vermont Advanced Computing Core,
  Department of Mathematics \& Statistics,
  The University of Vermont,
  Burlington, VT 05401.
  }

\author{
  \firstname{David Rushing}
  \surname{Dewhurst}
}

\affiliation{
  Vermont Complex Systems Center,
  Computational Story Lab,
  the Vermont Advanced Computing Core,
  Department of Mathematics \& Statistics,
  The University of Vermont,
  Burlington, VT 05401.
  }

\author{
  \firstname{Fletcher F.}
  \surname{Hazlehurst}
}

\affiliation{
  Vermont Complex Systems Center,
  Computational Story Lab,
  the Vermont Advanced Computing Core,
  Department of Mathematics \& Statistics,
  The University of Vermont,
  Burlington, VT 05401.
  }

\author{
  \firstname{Colin M.}
  \surname{Van Oort}
}

\affiliation{
  Vermont Complex Systems Center,
  Computational Story Lab,
  the Vermont Advanced Computing Core,
  Department of Mathematics \& Statistics,
  The University of Vermont,
  Burlington, VT 05401.
  }

\author{
  \firstname{Lewis}
  \surname{Mitchell}
}

\affiliation{
  School of Mathematical Sciences,
  North Terrace Campus,
  The University of Adelaide,
  SA 5005, Australia
}

\author{
  \firstname{Andrew J.}
  \surname{Reagan}
}

\affiliation{
  Vermont Complex Systems Center,
  Computational Story Lab,
  the Vermont Advanced Computing Core,
  Department of Mathematics \& Statistics,
  The University of Vermont,
  Burlington, VT 05401.
  }

\author{
  \firstname{Jake Ryland}
  \surname{Williams}
}

\affiliation{
  Department of Information Science,
  Drexel University,
  3141 Chestnut Street,
  Philadelphia, PA 19104.
  }

\author{
  \firstname{Christopher M.}
  \surname{Danforth}
}

\affiliation{
  Vermont Complex Systems Center,
  Computational Story Lab,
  the Vermont Advanced Computing Core,
  Department of Mathematics \& Statistics,
  The University of Vermont,
  Burlington, VT 05401.
  }

\date{\today}

\begin{abstract}
  \protect
  Herbert Simon's classic rich-get-richer model is
one of the simplest empirically supported mechanisms
capable of generating heavy-tail size distributions for complex systems.
Simon argued analytically that a population of flavored
elements growing by either adding a novel element or randomly
replicating an existing one
would afford a distribution of group sizes with a power-law tail.
Here, we show that, in fact, Simon's model
does not produce a simple power law 
size distribution as the initial element
has a dominant first-mover advantage,
and will be overrepresented by a factor proportional
to the inverse of the innovation probability.
The first group's size discrepancy cannot be explained 
away as a transient of the model, 
and may therefore be many orders of
magnitude greater than expected.
We demonstrate how Simon's analysis was correct but incomplete, 
and expand our alternate analysis to
quantify the variability of long term rankings for all groups.
We find that the expected time for a first replication is infinite,
and show how an incipient
group must break the mechanism to improve their odds of success.
We present an example of citation counts for a specific field
that demonstrates a first-mover advantage consistent
with our revised view of the rich-get-richer mechanism.
Our findings call for a reexamination of preceding work 
invoking Simon's model and provide an expanded understanding going forward.

\end{abstract}


\maketitle

\section{Introduction}
\label{sec:rgrfirstmover.introduction}

Across the spectrum of natural and constructed phenomena, 
descriptions of the architecture and dynamical behavior of complex systems 
repeatedly involve heavy-tailed distributions.
For systems involving components of variable size $S$,
many bear size distributions with power-law decays of the form 
$\Prob{\rgrSize} \sim \rgrSize^{-\gamma}$~\cite{simon1955a,newman2005b}:
word usage frequency in language~\cite{zipf1949a,ferrericancho2001c,williams2015b},
the number of species per genus~\cite{yule1925a,simon1955a},
citation numbers for scientific papers~\cite{price1965a,price1976a},
node degree in networks~\cite{barabasi1999a,krapivsky2001a,bornholdt2001a,newman2003a},
firm sizes~\cite{axtell2001a},
and the extent of system failures such as forest
fires~\cite{doyle2000a,carlson2002a}.
These size distributions are often alternately 
cast in the form of a Zipf distribution~\cite{zipf1949a}
with components ordered by decreasing size and
$\rgrSize_\zipfrank \sim \zipfrank^{-\zipfexponent}$
where $\zipfrank$ ($= 1, 2, \ldots$) is the size rank
and $\zipfexponent = 1/(\gamma-1)$~\footnote{
  We connect a power law size distribution
$
\prob{S}
\sim
\rgrSize^{-\gamma}
$
to the corresponding Zipf's law
$
\rgrSize_{\zipfrank}
\sim
r^{-\zipfexponent}
$
through the complementary cumulative
distribution
$
\probccdf{\rgrSize}
=
\int_{s=\rgrSize}^{\infty}
\dee{s}
\prob{s}
\sim
S^{-\gamma+1}.
$
For a finite system with $N$ elements,
$
N
\probccdf{\rgrSize}
$
estimates the number of groups with
size at least $\rgrSize$.
We see that this must also be the rank $\zipfrank$
of the group of size $\rgrSize_{\zipfrank}$,
and we have
$
\zipfrank
\simeq
N
\probccdf{\rgrSize_{\zipfrank}}
\sim
N
\rgrSize_{\zipfrank}^{-\gamma+1}.
$
Because
$
\rgrSize_{\zipfrank}
\sim
r^{-\zipfexponent},
$
we therefore have
$\gamma
=
1
+
1/\zipfexponent$
and equivalently
$
\zipfexponent
=
1/(\gamma-1)
$
  }.

Elucidating and understanding the most essential dynamical models 
leading to power-law size distributions is an essential task.
While the mechanisms giving rise to such distributions 
are diverse, they generally involve growth and replication. 

In his famous 1955 paper on skewed distributions~\cite{simon1955a},
Simon built on classical urn model theory
to show that a simple, single parameter, rich-get-richer mechanism
could lead a growing population to produce
a pure power-law size distribution of groups
of elements of matching type~\cite{wikipedia-yule-simon-distribution2017a}.
Simon's model
is governed by an innovation probability $\innovationrate$
which Simon argued controls the 
group size distribution exponent 
as 
$\gamma = 1 + 1/(1-\innovationrate)$,
and, equivalently,
the Zipf's law exponent as
$\zipfexponent = 1 - \innovationrate$
(we re-derive these results as part
of our analysis in Sec.~\ref{sec:rgrfirstmover.analysis}).

Simon's model has endured because it is at once a boiled-down,
easy-to-understand toy model
representative of a large class of rich-get-richer mechanisms,
and yet it is also a model that has a remarkable ability to capture the essential
growth dynamics of disparate, real-world complex systems.
While not without controversy, particularly for language~\cite{mandelbrot1953a,mandelbrot1959a,simon1960a,mandelbrot1961a,simon1961a,mandelbrot1961b,simon1961b,miller1965a,ferrericancho2010a},
Simon's micro-to-macro link between the separately measurable
innovation rate and power-law scaling for system component
size distribution has been observed to roughly hold
for
word counts in books~\cite{simon1955a},
citation counts in scientific
literature~\cite{krapivsky2001a,newman2009b,newman2014a},
the early growth of the Web~\cite{bornholdt2001a},
and
the development of software such as the Linux kernel~\cite{maillart2008a}.

rich-get-richer models adjacent to Simon's model
have been employed to characterize the essential features
of many kinds of systems such as the
emergence of novelties~\cite{tria2014a,loreto2016a}.
Arguably the most profound role of rich-get-richer mechanisms
has been uncovered in complex networks.
Simon's model is the explicit
core of Price's cumulative advantage
mechanism for the growth of citation networks in scientific
literature~\cite{price1965a,price1976a}.
A modified version of Simon's model is also at the heart of
the independently discovered growing network model
of preferential attachment due to the field-starting work
of Barab{\'a}si and Albert~\cite{barabasi1999a}.

Here, we show analytically and through simulations that Simon's
analysis, for all its successes, was strikingly incomplete:
The initial group enjoys
a profound ``first-mover advantage'' on the order of the inverse
of the innovation probability, $1/\innovationrate$.
This is not a small correction to a long established theory.
As the innovation probability is typically less than 0.1 and
often much closer to 0~\cite{zipf1949a,bornholdt2001a,newman2005b,maillart2008a,williams2015a,williams2015b},
the initial group's size may be orders of magnitude
greater than would be consistent with a simple power law.
Nor, as we will show, can the first group be dismissed as a transient or
as a kind of null group and not part of the system.
Indeed, we provide evidence from scientific citation data
that a first mover advantage manifested by Simon's model
is a real phenomenon.

In what follows, we first describe Simon's model
and
present results from simulations (Sec.~\ref{sec:rgrfirstmover.simulations});
analytically determine the first-mover's advantage (Sec.~\ref{sec:rgrfirstmover.analysis});
explore the detailed long term dynamics of all groups (Sec.~\ref{sec:rgrfirstmover.variability});
self-referentially compare the model's output with citation data
concerning scale-free networks~\cite{barabasi1999a} (Sec.~\ref{sec:rgrfirstmover.evidence});
and, finally, 
consider broader implications for understanding 
real-world, rich-get-richer systems
(Sec.~\ref{sec:rgrfirstmover.conclusion}).

\begin{figure*}
  \centering
              \includegraphics[width=\textwidth]{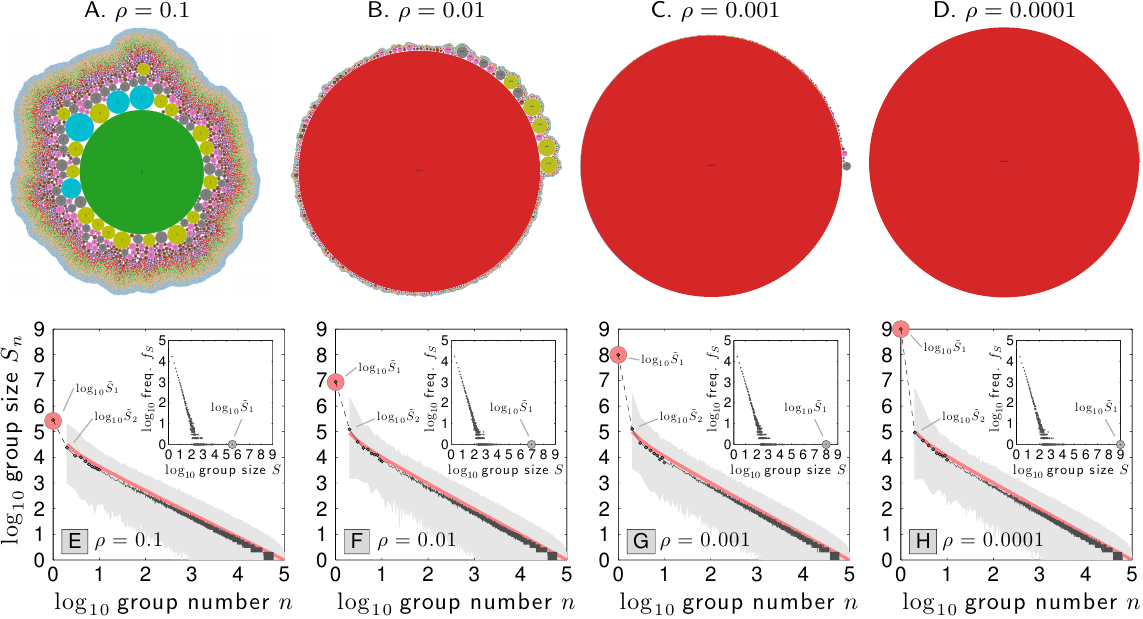}
  \caption{
    Results from simulations of Simon's model showing 
    an inherent dominant first-mover advantage.
    \textbf{A--D.}
    Visualizations of group sizes after approximately
    $t=10^6$ time steps
    for 
    $\innovationrate$ = 0.1, 0.01, 0.001, and 0.0001.
    Group sizes are proportional to disk area.
    The colors are a function of each simulation's specific history
    but do match across groups of the same size within each system.
    \textbf{E--H.}
    Main plots: Zipf-like distributions for an ensemble of
    200 simulations in gray and theory in red [see \Req{eq:rgrfirstmover.Nsimoncorrected}].
    Differing from the simulations for the visualizations,
    these were run until $\groupnumt{} = 10^5$, 
    approximately $10^5/\innovationrate$ time steps.
    The dark gray curve indicates the median size
    of the $\groupnum$th arriving group and the light gray bounds
    2.5 to 97.5 percentiles.
    For clarity, the median sizes of the first and second groups
    are highlighted 
    as
    $\log_{10} \tilde{S}_1$
    and
    $\log_{10} \tilde{S}_2$.
    Insets:
    For each value of $\innovationrate$,
    we show the (raw) size frequency distribution
    $f_\rgrSize$
    for a single simulation,
    again indicating $\log_{10} \tilde{S}_1$.
    The exponents for the Zipf distribution and
    size distributions
    are connected as $\gamma = 1 + 1/\zipfexponent$.
    We provide interactive simulations for Simon's model
    as part of the paper's online appendices~\cite{dodds2016b_onlineappendices}.
  }
  \label{fig:rgrfirstmover.zipfdists}
\end{figure*}

\section{Evidence of a first mover advantage from simulations}
\label{sec:rgrfirstmover.simulations}

The algorithm for Simon's classic rich-get-richer model is simple.
We are concerned with the growth of a population of elements 
where each element has a type, and elements of the same
type form a group.
In modeling real systems, types may represent a city, a word,
or the destination of a link in a network~\cite{simon1955a,maillart2008a}.
Beginning with a single element at time $t=1$,
an element is added to the population
at each discrete time step $t \ge 2$.
Representing innovation, the arriving element
has, 
with probability $\innovationrate$,
a new, previously unseen type.
Alternately, with probability $1-\innovationrate$,
the arriving element is a replication taking
on the type of a randomly chosen existing element.
As one of a number of generalizations,
we will,
as Simon did himself in~\cite{simon1955a},
later consider
a dynamic innovation probability $\innovationratet$
in Sec.~\ref{sec:rgrfirstmover.conclusion}.

In Fig.~\ref{fig:rgrfirstmover.zipfdists}, we provide
visualizations along with Zipf-like distributions
and size distributions 
for Simon's model for 
$\innovationrate$ = 0.1, 0.01, 0.001, and 0.0001.
In Figs.~\ref{fig:rgrfirstmover.zipfdists}A--D, 
we show results for four sample simulations corresponding to
these values of $\innovationrate$ after $10^6$ time steps.
Each disk represents a group with area proportional to group size.
Interactive simulations from which these images are drawn
are available in the paper's
online appendices~\cite{dodds2016b_onlineappendices}.

To the eye, for $\innovationrate=0.1$, the first group appears to 
be somewhat outsized but perhaps not inconsistent.
However, for the next three (decreasing) values of $\innovationrate$,
the first group is evidently different, increasingly accruing
the bulk of all new elements.

In Figs.~\ref{fig:rgrfirstmover.zipfdists}E--H, we make clear the
first-mover advantage through
Zipf-like distributions for group size (main plots) and size distributions
(inset plots)
for the same ordering of $\innovationrate$ values.
The group sizes in the main plots are a function of group arrival
number
rather than rank according to decreasing size (hence Zipf-like),
and 
come from an ensemble of simulations (median in dark gray, 2.5 to 97.5
percentile range in light gray; see caption for details)
and overlaid theory (red).
We examine the variation in group size with group arrival
order later on.
As per Simon's analysis,  median group size $\tilde{\rgrSize}_{n}$ behaves as $n^{-\zipfexponent}$ for $n \ge 10$
where 
$\zipfexponent$ $(=1-\innovationrate)$ 
is the Zipf exponent.
However, as we demonstrate theoretically below,
the dominant group $\tilde{\rgrSize}_{1}$ is larger than would be expected
by a factor of $1/\innovationrate$.

Again for single example simulations, the inset size frequency distributions
(raw counts; notation: $f_{\rgrSize}$)
in Figs.~\ref{fig:rgrfirstmover.zipfdists}E--F
show the same disparity
with the largest element isolated from the 
main power-law-obeying size distribution.
We can now see why Simon's analysis,
while technically correct, fell short:
The size distribution he derived fit all but one point which,
if not observed and handled appropriately,
vanishes in contribution in the infinite system size limit.

\section{Analytic determination of the first-mover advantage}
\label{sec:rgrfirstmover.analysis}

To understand this first-mover advantage,
we carry out a re-analysis of Simon's model.
New groups are initiated
stochastically
with the $\groupnum$th group first appearing on average at time
$\grouptimestart{\groupnum}$.
We write the number of elements in the $\groupnum$th group
at time $t \ge \grouptimestart{\groupnum}$ as $\groupsizet{\groupnum}{t}$,
and each group starts with a single element:
$\groupsizet{\groupnum}{\grouptimestart{\groupnum}} = 1$
(we examine this choice later).
As Simon did, we assume an initial condition
of a population of 1 element at time $t=1$
(we consider a general initial condition
later in Sec.~\ref{sec:rgrfirstmover.conclusion},
\Req{eq:rgrfirstmover.Nsimoncorrected_initialseed}).
If at time $t$, we have $\groupnumt{t}$ distinct groups,
the probability that a randomly drawn 
element belongs to the $\groupnum$th group is then:
\begin{equation}
  \groupprobt{\groupnum}{t}
  =
  \frac{\groupsizet{\groupnum}{t}}
  {\sum_{\groupnum'=1}^{\groupnumt{t}} \groupsizet{\groupnum'}{t}}
  =
  \frac{\groupsizet{\groupnum}{t}}
  {t}.
  \label{eq:rgrfirstmover.probability}
\end{equation}

We construct an evolution equation for the size of 
the $\groupnum$th arriving group, $\groupsizet{\groupnum}{t}$.
At time $t$, for the $\groupnum$th group to increase in number by 1,
replication must be chosen 
(occurring with probability $1-\innovationrate$),
and then an element in
the $\groupnum$th group must be replicated, 
leading to the probablistic statement:
\begin{equation}
  \avg{
    \groupsizet{\groupnum}{t+1}
    -
    \groupsizet{\groupnum}{t}
  }
  =
  (1-\innovationratet)
  \cdot
  \frac{
    \groupsizet{\groupnum}{t}
  }
  {
    t
  }
  \cdot
  (+1).
  \label{eq:rgrfirstmover.Nt}
\end{equation}
In the case of Simon's fundamental model with a fixed
innovation probability,
$\innovationratet = \innovationrate$,
we proceed
with a difference equation calculation.
Our primary analysis is approximate
as we drop the expectation on the left hand side
of~\Req{eq:rgrfirstmover.Nt}.

For fixed $\innovationratet = \innovationrate$ and shifting from $t$ to $t-1$,
\Req{eq:rgrfirstmover.Nt} 
gives the approximation, 
for $t \ge \grouptimestart{\groupnum}$,
\begin{equation}
  \groupsizet{\groupnum}{t}
  =
  \left[
    1
    +
    \frac{(1-\innovationrate)}
    {t-1}
  \right]
  \groupsizet{\groupnum}{t-1}.
  \label{eq:rgrfirstmover.Ntdiff}
\end{equation}
Again given that 
$
\groupsizet{\groupnum}{\grouptimestart{\groupnum}}
=
1
$,
we have
\begin{gather}
  \groupsizet{\groupnum}{t}
  =
  \left[
    1
    +
    \frac{(1-\innovationrate)}
    {t-1}
  \right]
  \left[
    1
    +
    \frac{(1-\innovationrate)}
    {t-2}
  \right]
  \cdots
  \left[
    1
    +
    \frac{(1-\innovationrate)}
    {\grouptimestart{\groupnum}}
  \right]
  \cdot
  1
  \nonumber
  \\
  =
  \left[
    \frac{t - \innovationrate}
         {t-1}
         \right]
  \left[
    \frac{t -1 - \innovationrate}
         {t-2}
         \right]
  \cdots
  \left[
    \frac{\grouptimestart{\groupnum} + 1-\innovationrate}
    {\grouptimestart{\groupnum}}
  \right]
  \nonumber
  \\
  =
  \frac{
    \Gamma(t+1-\innovationrate) 
    \Gamma(\grouptimestart{\groupnum})
  }
  {
    \Gamma(\grouptimestart{\groupnum}+1-\innovationrate) 
    \Gamma(t)
  }
  =
  \frac{
    \Beta(\grouptimestart{\groupnum},1-\innovationrate)
  }
  {
    \Beta(t,1-\innovationrate)
  },
  \label{eq:rgrfirstmover.Ntdiffevolve}
\end{gather}
where $B$ is the beta function.
Our analysis thus far has been heading towards the 
same conclusion as Simon's which took
a different route following the
evolution of the system's size distribution.
Now, however, we find a distinction that renders
the initial group special.
For constant $\innovationrate \ll 1$,
the $\groupnum$th group first arrives, on average,
at 
$
\grouptimestart{\groupnum} 
\simeq
[
\frac{\groupnum-1}
{\innovationrate}
]
$
where 
$[ \, \cdot \, ]$
is the rounding operator.
But this is only valid for $\groupnum \ge 2$ 
because
$
\grouptimestart{1} = 1.
$
For large $t$,
we find the size of the $\groupnum$th 
arriving group to be:
\begin{equation}
  \groupsizet{\groupnum}{t}
  \sim
  \left\{
    \begin{array}{l}
        \frac{1}
        {\Gamma(2 - \innovationrate)}
      \left[
        \frac{1}
        {t}
      \right]^{-(1-\innovationrate)}
      \
      \textnormal{for}
      \
      \groupnum = 1,
      \\
      \innovationrate^{1 - \innovationrate}
      \left[
        \frac{\groupnum - 1}
        {t}
      \right]^{-(1-\innovationrate)}
      \
      \textnormal{for}
      \
      \groupnum \ge 2.
      \\
    \end{array}
  \right.
  \label{eq:rgrfirstmover.Nsimoncorrected}
\end{equation}
where we have used
the asymptotic scaling 
$\Beta(x,y) \sim \Gamma(y) x^{-y}$ for large $x$,
and replaced 
$t + 1 - \innovationrate$ with $t$.

On average, we expect the groups to be ranked by decreasing
size according to their arrival number $\groupnum$.
For large $\groupnum$, 
Zipf's law appears with 
$
\groupsizet{\groupnum}{t}
\sim
\innovationrate^{1 - \innovationrate}
\left[
  \frac{\groupnum}
 {t}
\right]^{-\zipfexponent}
$
with exponent 
$\zipfexponent = 1 - \innovationrate$.
As we move down toward $\groupnum = 2$,
a small correction factor arises as
$
\left[
  \frac{\groupnum - 1}
  {\groupnum}
\right]^{-\zipfexponent}
$,
maximally $2^{1-\rho} < 2$ at $\groupnum = 2$.

We now see that the initial and largest group manifestly does not conform
because of the absence of the $\innovationrate^{(1-\innovationrate)}$ term.
Moreover, as $\innovationrate \rightarrow 0$,
the size of the dominant group departs rapidly from Zipf's law
with a factor of $1 \cdot \innovationrate^{-1}$
greater than what would be internally consistent with 
$\groupsizet{\groupnum}{t}$ 
for 
$\groupnum \gg 2$.

Shifting to the size frequency distribution~\cite{Note1},
we have
$
\Prob{\rgrSize}
\propto
\rgrSize^{-\gamma} 
= 
\rgrSize^{
  -\left(
  1+
  \frac{1}
       {\zipfexponent}
       \right)
}
= 
\rgrSize^{
  -\left(
  1+
  \frac{1}
    {1-\innovationrate}
  \right)
}
$,
and again for values of $\innovationrate$ consistent with
real-world data~\cite{simon1955a,newman2005b},
the first group is an outlier clearly separated
from the power-law size distribution.

Simon's analysis,
like many subsequent treatments~\cite{krapivsky2001a,newman2009b,newman2014a,hebert-dufresne2012a,berset2013a,fotouhi2013a,gleeson2014a,gleeson2016a},
was a mixture
of discrete and continuous
pieces,
obtaining the
tail of the group size distribution through an asymptotic expansion.
Such a continuum approximation washed out the discrete nature
of the system rendering the first mover anomaly hidden.
Further, because the first mover is only one of an infinite number
of groups for $t \rightarrow \infty$, its weight in the distribution
tends to 0 and will thus be unobserved.

While it may be tempting to interpret the first group
as a so-called ``dragon-king''~\cite{sornette2012a,wikipedia-king-effect2017a},
we see that it is an endogenously generated product
of an elementary rich-get-richer model rather than
an exogenous mechanism singularly affecting the largest events
within a system.

The preferential attachment models developed in~\cite{hebert-dufresne2012a}
show evidence of a first mover advantage and may profit
from an analysis similar to the one we have laid out here.

\begin{figure*}
  \centering
  \includegraphics[width=\textwidth]{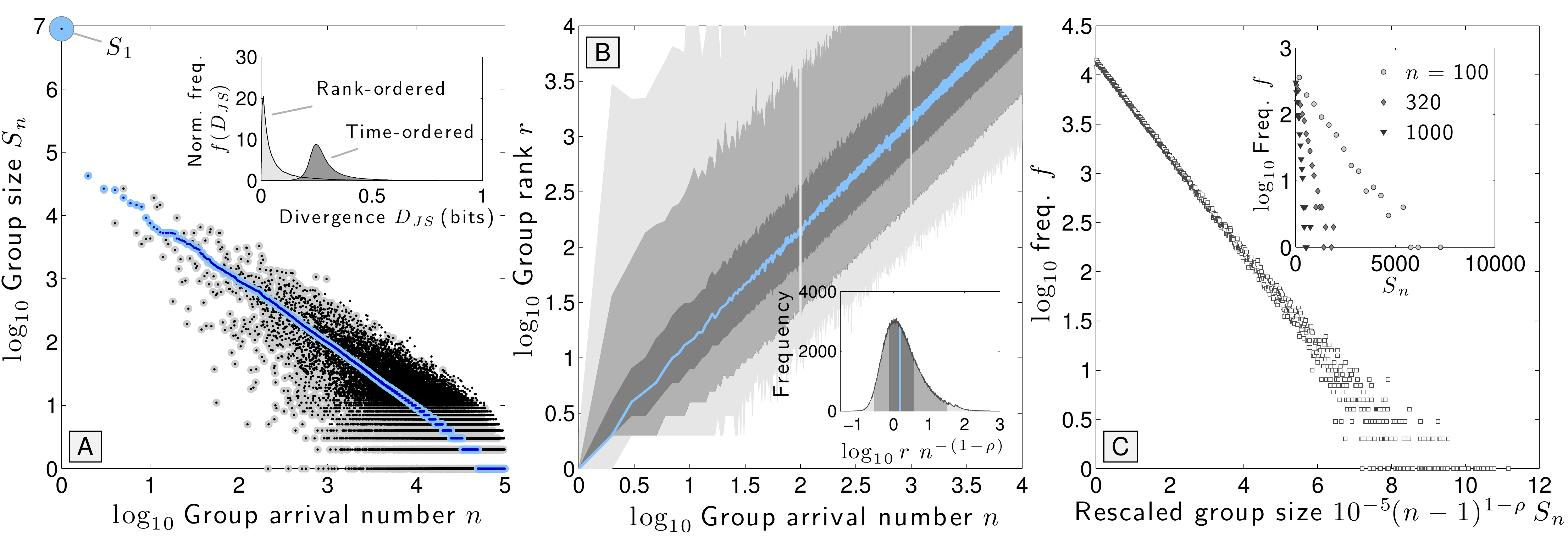}
  \caption{
    Simulation results for a numerical 
    investigation of the variability of final rank $\zipfrank$ as a function of group
    arrival number $\groupnum$. 
    All panels are derived from an ensemble of 1000 systems with
    $\innovationrate$ = 0.01 evolved
    for $10^7$ time steps.
    \textbf{A.}
    Main panel:
    For a single example system, group size as a function of 
    group arrival number $\groupnum$ (black)
    with ranked sizes overlaid (blue).
    Inset:
    Comparison of Jenson-Shannon divergences for all pairs of
    group sizes as a function of time of arrival (time-ordered)
    and all pairs of groups sizes when ranked by size (rank-ordered).
    \textbf{B.}
    Main panel:
    Group rank $\zipfrank$ as a function of group arrival number
    $\groupnum$
    with median in light blue and 
    the increasingly gray regions delimiting minimum-maximum, 
    2.5 to 97.5 percentiles,
    and 
    25 to 75 percentiles.
    Inset:
    Rescaled distribution of
    group ranks for $10^2 \le \groupnum \le 10^3$ (white lines in main plot).
    \textbf{C.}
    Main panel:
    Rescaled 
    size distributions for $10^2 \le \groupnum \le 10^3$
    indicating an exponential form.
    Inset:
    Raw size distributions for three example arrival group numbers,
    $\groupnum$ = 100, 320, and 1000.
  }
  \label{fig:rgrfirstmover.richgetricher_paper_ordering_timescale003}
\end{figure*}

\section{Long term variability of group success}
\label{sec:rgrfirstmover.variability}

We now delve into how arrival order of groups 
($\groupnum$)
relates to final rank
($\zipfrank$),
an issue arising with the Zipf-like distributions we presented
in Fig.~\ref{fig:rgrfirstmover.zipfdists} and elided in our preceding
analysis.

While it is evident that Simon's model must produce a degree
of indeterminacy in ranking, our goal here is largely to use
simulations and some specific analysis to characterize the
nature and extent of rank variability
for growing network models based on Simon's model~\cite{price1976a,barabasi1999a}.
Krapivsky and Redner \cite{krapivsky2001a}
and Newman~\cite{newman2009b,newman2014a} have analytically determined
the distribution of citations that the $\groupnum$th arriving group
will receive as a function of time.
Further, Newman was able to show good agreement with citation
data.
Our focus on the original Simon model and different treatment
means we arrive at some complementary findings.

In the main plot of Fig.~\ref{fig:rgrfirstmover.richgetricher_paper_ordering_timescale003}A,
we show results for a single simulation 
with
$\innovationrate = 0.01$,
and $10^7$ total elements (equivalent to run time),
and
$\groupnumt{} \simeq 10^5$ distinct groups.
The dark points indicate the actual
size of the $n$th group while the blue curve is
the resulting, properly ranked Zipf distribution.
While the first mover dominates as expected,
and the second arriving group obtains
the second ranking in this instance,
we observe substantial and increasing decoupling
between arrival order and final rank as the
system grows.

To further explore this decoherence, we generate an ensemble of 1000
simulations of the same system.
The effect on final group ordering brought about by decreasing $\innovationrate$ saturates quickly
and all of the following results regarding arrival time and rank
are essentially the same for $\innovationrate \le 0.01$ with little quantitative change.
We first measure the variability of the resulting distributions using the Jenson-Shannon
divergence (JSD), time-ordered against time-ordered and rank-ordered
against rank-ordered, a total of $\binom{1000}{2}$ such comparisons
for each.
In the inset to Fig.~\ref{fig:rgrfirstmover.richgetricher_paper_ordering_timescale003}A, we show the
distributions of JSD for each ordering, finding a typical
disparity for the time-ordered size distributions of 0.25 bits.

In the main plot of
Fig.~\ref{fig:rgrfirstmover.richgetricher_paper_ordering_timescale003}B, we
present 
overall group rank $\zipfrank$
as
a function of group arrival number 
$\groupnum$
for our ensemble of 1000
simulations.
The pale blue line indicates the median,
and the surrounding gray regions mark
the 25 to 75 percentile range,
the 2.5 to 97.5 percentile range,
and the minimum to maximum range.
In the inset, we rescale and collapse the final rank distribution
for arrival group numbers $10^2$ to $10^3$
based on \Req{eq:rgrfirstmover.Nsimoncorrected}.
We see that from around just the 10th arriving group on,
95\% of a group's final rank spans 
a remarkable two orders of magnitude
around the median $r = n$, skewed towards higher values 
as shown in the inset.
Thus, while by equating rank and arrival number,
our analysis is effective for the median size
of groups, 
the system's specific dynamics are considerably more complex.

Lastly, in Fig.~\ref{fig:rgrfirstmover.richgetricher_paper_ordering_timescale003}C, we examine
the distribution of possible group sizes as a function of group 
arrival number $n$.
Rather than possessing a single maximum that increases with $n$,
we find that an exponential distribution is a good approximation.
The inset gives three example distributions and the main
plot shows an appropriate rescaling of all $100 \le n \le 1000$.

For all but a small initial collection of groups, the mode group size
is thus close to 1, 
consistent with Simon's asymptotic result that
$1/(2-\innovationrate) 
\simeq 
1/2$
of all groups contain only one element (hapax legomena for texts).
To understand how half of all initiated groups never grow in size,
we determine the probability of a group gaining a new member
as a function of its arrival time and current size.
The probability that the $\groupnum$th arriving group,
which is of, say, size 
$\groupsizet{\groupnum}{t}=k$
at time $t$,
fails to replicate for all of times $t$ through $t+\tau-1$ before
replicating at time $t + \tau$ is:
\begin{gather}
  \Prob{
    \groupsizet{\groupnum}{t+\tau}
    =
    k+1 
    \, 
    \middle|
    \, 
    \groupsizet{\groupnum}{t+i} 
    = k
    \,
    \mbox{for $i=0, \ldots, \tau-1$}
  }
  \nonumber
  \\
  =
  \prod_{i=0}^{\tau-1}
  \left[
    1
    -
    {(1-\innovationrate)}
    \frac{k}
    {t+i}
  \right]
  \cdot
  {(1-\innovationrate)}
  \frac{k}
  {t+\tau}
  \nonumber
  \\
  =
  k
  \frac{
    B(\tau,t)
  }
  {
    B\left(
      \tau,t-(1-\innovationrate)
      \right)
  }
  \frac{1-\innovationrate}
  {t+\tau}
  \propto
  \frac{\tau^{-(1-\innovationrate)k}}
  {t + \tau},
  \label{eq:rgrfirstmover.probgrowth}
\end{gather}
where we have again used
that $\Beta(x,y) \sim \Gamma(y) x^{-y}$ for large $x$.
We observe a power-law decay with two scaling regimes.
For $\tau \gg t$, the probability behaves as
$\tau^{-1-(1-\innovationrate)k}$.
As the $\groupnum$th arriving group 
starts with $\groupsizet{\groupnum}{\grouptimestart{\groupnum}} = 1$ element,
the exponent is $-(2-\innovationrate) > -2$
and the expected time for replication is infinite.
Once a group has replicated, the expected time becomes finite.
A newly arriving group is therefore greatly advantaged 
if it can step out of the mechanism and begin with even
just two elements.
Another variation allowing for later success is
for elements to have variable inherent qualities
as has been done, for example, for the Barab{\'a}si-Albert model~\cite{barabasi1999a,bianconi2001a,bianconi2001b}.

\section{Empirical evidence of a first mover advantage obeying Simon's model}
\label{sec:rgrfirstmover.evidence}

\begin{figure}
  \centering
  \includegraphics[width=0.48\textwidth]{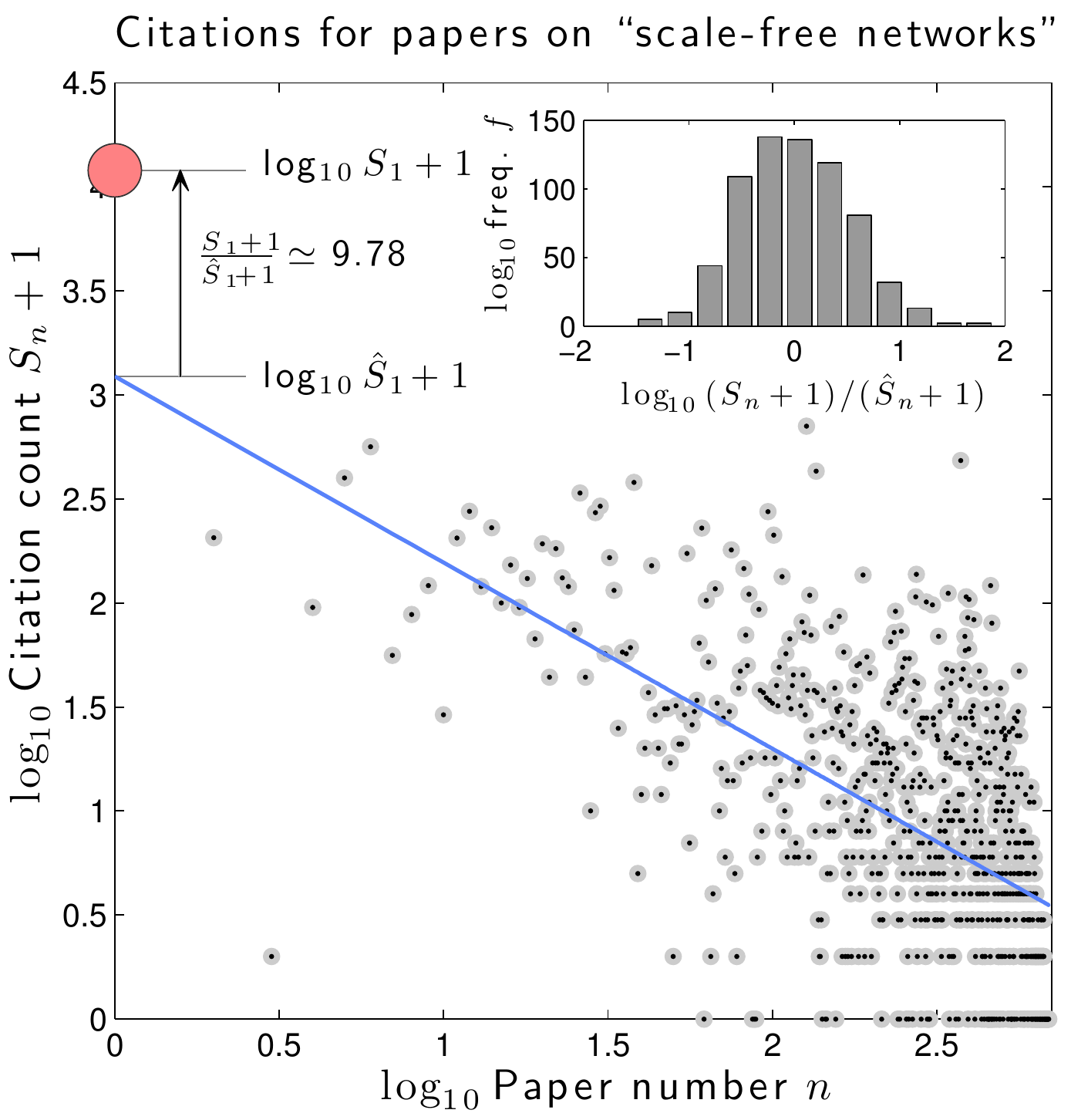}  
  \caption{
    Citation counts for (1) the original
    paper on scale-free networks by
    Barab{\'a}si and Albert~\cite{barabasi1999a} (larger pale red disk)
    and (2)
    692 subsequent papers with ``scale-free networks'' in their
    title ordered by publication date.
    We obtained citation counts from the Web of
    Science~\cite{webofscience2016a} on October 16, 2016.
        The blue line is the line of best fit using standard
    linear regression for all articles
    excluding the initiating article.
    A slope of $-0.89 \pm 0.09$
    gives an estimate of $\innovationrate \simeq  0.11 \pm 0.09$.
    The first-mover is a factor of 9.78 greater than
    would be consistent with the line of best fit,
    approximately the expected ratio of
    $1/\innovationrate = 1/0.11 = 9.46$.
    The inset show the binned residuals span two orders
    of magnitude, in keeping with the results shown
    in the inset of Fig.~\ref{fig:rgrfirstmover.richgetricher_paper_ordering_timescale003}B.
  }
  \label{fig:rgrfirstmover.webofscience_scalefreenetworks_zipf003}
\end{figure}

While we do not expect systems with
truly pure Simon mechanisms operating
from a single initial element to be widespread
(more below),
one possible application lies in citation data for
papers central to the development of a well-defined area of research.
In
Fig.~\ref{fig:rgrfirstmover.webofscience_scalefreenetworks_zipf003},
we show citation counts for papers with titles
matching ``scale-free networks'' (gray dots) \footnote{
  We have chosen the restriction to such titles so as to limit
  ourselves to publications that are primarily concerned
  with the topic of scale-free networks;
  we acknowledge the crude nature of this approach
  and that better and far more exhaustive analyses
  would be available for future work.
}
along with citations counts for the incipient work of
Barab{\'a}si and Albert~\cite{barabasi1999a}
(large red disk) (in the manner of~\cite{price1976a},
we add 1 to all citation counts).
We used Web of Science~\cite{webofscience2016a} to
find these papers and their citation counts
obtaining a total of 692 on October 16, 2016.
We note that although Barab{\'a}si and Albert
introduced the term in~\cite{barabasi1999a},
they did not elevate it to being
part of their own title.
Of course, our model is not a perfect fit structurally
as the citations are from all sources and
each new paper would likely cite those coming before,
especially~\cite{barabasi1999a};
we are also not attempting to 
capture the growth of a citation network.

Because our results show that a pure Simon model will produce
a Zipf distribution with exponent $\zipfexponent = 1 - \innovationrate$
and a first-mover factor of $1/\innovationrate$,
we have two distinct measurable quantities that
estimate $\innovationrate$.
Regressing through all but the initial paper's citations counts (blue line),
we obtain a Zipf exponent $\zipfexponent = 0.89 \pm 0.09$ which in
turn
gives the estimate
$\hat{\innovationrate} = 1 - \zipfexponent = 0.11 \pm 0.09$.
The ratio of the number of citations for the first paper
relative to that indicated by the regression is 9.78 which
gives a very comparable value of $\hat{\innovationrate} = 1/9.78
\simeq 0.10$,
one that is well within standard error of the first estimate.

The inset
for Fig.~\ref{fig:rgrfirstmover.webofscience_scalefreenetworks_zipf003}
shows the binned residuals for citation counts around
the line of best fit for Zipf's law.
The two orders of magnitude variation in citation counts
conforms well with results from simulations displayed
in the inset of Fig.~\ref{fig:rgrfirstmover.richgetricher_paper_ordering_timescale003}B.

Presuming all 692 papers cite the original
Barab{\'a}si and Albert paper,
then
a fraction 692/11982 $\simeq$ 0.058
of the first paper's citations
come from papers centrally occupied
with scale-free networks.
Removing these and all citations internal
to the set of 693 papers would not greatly
alter the observed congruence between
the first mover's outsized citation count
and Zipf's law.

Improving and extending this analysis of citation data
would be natural if potentially difficult to automate at a large scale.
To be a potential fit for the pure Simon's model with the first mover advantage
considered a feature,
a new research area would need to be started cleanly with a single
paper.
The paper would have to introduce a singular and lasting catchphrase for the
focal topic, one that subsequent researchers developing the area would see fit to be included
in a paper title.
In order to build a sufficiently large enough data set,
the analysis may also be successfully expanded to include papers
with matches for the catchphrase in their abstracts rather than
titles.
We also note that we have ignored real time in our present analysis
and that some modification of the model will be
needed to properly accommodate the probability that
the $\groupnum$th relevant paper appears at time $t$.

\section{Concluding remarks}
\label{sec:rgrfirstmover.conclusion}

We close with some thoughts on how the dominant first-mover
advantage of Simon's model
and the variability of ordering 
may confound empirical analyses of real-world systems,
and offer some potential resolutions.

As an idealized process, Simon's model evidently and purposefully
fails to involve many aspects and details of real world systems.
Nevertheless, we must address the issue that Simon's rich-get-richer model 
has performed extremely well in analyses of real systems that do not exhibit a dominant group.
Particular successes~\cite{williams2016a} have been found in measures of
the innovation probability $\innovationrate$
and 
the Zipf exponent $\zipfexponent$ with
the expectation $\zipfexponent = 1-\innovationrate$,
as well as in the fraction of groups with one or two
members~\cite{simon1955a,price1976a,bornholdt2001a,maillart2008a}

In moving away from a pure Simon model,
we consider three variant conceptions.
A first possibility is to treat the first group as being
of a different kind to those that come after
by declaring the first element to be
a kind of unobservable null element---its selection and replication
at time $t$ represents a failure of the system to produce
any visible element.
Given our analysis above, Zipf's law would shift 
with $\groupnum \rightarrow \groupnum - 1$ and then be
closely approximated by a power law decay~\cite{williams2016a}.
However, this is a problematic mechanism
as it requires the null element to be hidden from an
observer of the system,
while at the same time visible and equal in nature to all other
elements from the point of view of the replication mechanism~\footnote{One possible exception may be language
  evolution, with whitespace framed as a kind of null element~\cite{williams2016a}.}.
This mechanism is not the same as
the one formed by adding an overall 
master update probability to Simon's model 
(i.e., at each time step, engage the element-adding mechanism
according to some fixed probability $\innovationrate_{\textnormal{add}}$).
Such a modification would only serve to slow the dynamics.
Further, starting the original model with no element at time $t=1$
would also fail for as soon as the first element appears,
the model would then act in the same way.
The first-mover advantage of Simon's model cannot be 
dismissed by any meaningful reinterpretation of the mechanism.

A second modification would be to allow the innovation probability
to vary with time (something Simon considered in~\cite{simon1955a};
see also~\cite{gerlach2013a,williams2015b}).
Such a dynamic $\innovationratet$ is plausible for real world
systems~\cite{baumol1986a}, and is exemplified by Heap's law,
the observation that ``word birth'' rate decays
as a function of text or corpus length~\cite{heaps1978a,herdan1960a,petersen2012b,pechenick2015b}.
For example, the innovation rate may be initially high with
new groups appearing rapidly,
while in the long run, the system may stabilize in its expansion rate
with the innovation probability $\innovationratet$ dropping
and tending towards a constant.
For such a dynamic, we would obtain a power-law tail for Zipf's law,
but early on, a high innovation probability would suppress
the first group, and smooth out the overall distribution.
While in principle $\innovationratet$ could be estimated from data,
great care would have to be taken given the stochastic evolution 
of a single run of a pure Simon model that we have demonstrated above
and observed in real world systems~\cite{bornholdt2001a,maillart2008a}.

Finally, and perhaps more realistically, we may have a system that is initially configured 
by an entirely different growth mechanism up until some
time $t_0$ at which a pure Simon model takes over.
Building on our earlier analysis, we can instantiate a simple version of such a system with
an initial $t=1$ condition of 
$\rgrSizeinit$ elements spread over
$\groupnuminit$ groups ranked by size
($\groupsizet{n}{1} \ge \groupsizet{n+1}{1}$ 
for 
$1 \le n \le \groupnuminit$).
Allowing the basic rich-get-richer mechanism to then go into effect,
the same approach of
Eqs.~\ref{eq:rgrfirstmover.Nt}--\ref{eq:rgrfirstmover.Nsimoncorrected}
returns
\begin{gather}
  \groupsizet{\groupnum}{t}
  =
  \frac{
    \Beta(\grouptimestart{\groupnum},1-\innovationrate)
  }
  {
    \Beta(\rgrSizeinit + t-1, 1-\innovationrate)
  }
  \groupsizet{n}{\grouptimestart{\groupnum}}.
\end{gather}
For large $t$, we have the approximate result:
\begin{gather}
  \groupsizet{\groupnum}{t}
  \sim
  \left\{
    \begin{array}{l}
        \frac{\groupsizet{n}{1}}
        {\Gamma(2 - \innovationrate)}
      \left[
        \frac{1}
        {\rgrSizeinit + t-1}
      \right]^{-(1-\innovationrate)}
            \
      \textnormal{for}
      \
      1 \le n \le \groupnuminit,
      \\
      \innovationrate^{1 - \innovationrate}
      \left[
        \frac{\groupnum - \groupnuminit}
        {\rgrSizeinit + t-1}
      \right]^{-(1-\innovationrate)}
      \
      \textnormal{for}
      \
      \groupnum > \groupnuminit,
      \\
    \end{array}
  \right.
  \label{eq:rgrfirstmover.Nsimoncorrected_initialseed}
\end{gather}
which reduces to \Req{eq:rgrfirstmover.Nsimoncorrected} 
if 
$
\groupnuminit = 
\rgrSizeinit = 
\groupsizet{1}{1} = 
1$.
\Req{eq:rgrfirstmover.Nsimoncorrected_initialseed}
shows how the first-mover advantage is distributed
across the groups present when Simon's model is introduced,
and quantifies how the likelihood for newly arriving groups to replicate
diminishes
as the population size of the initial elements increases.
For real-world systems with Zipf's laws that have clean
power-law tails leading out from difficult-to-characterize
forms for the largest groups, 
\Req{eq:rgrfirstmover.Nsimoncorrected_initialseed}
offers a possible fit that would also estimate
a time of transition from an establishing mechanism
to a rich-get-richer one.

A parallel finding is that the initial network structure also matters for
the Barab{\'a}si-Albert growing network
model~\cite{berset2013a,fotouhi2013a}, offering the
possibility that the initial conditions and onset of the preferential
attachment mechanism may be estimated from an observed network.
For even relatively small seed networks with average
degree 10, the asymptotic behavior of the degree distribution
will not match the classic $k^{-3}$ form found
in~\cite{barabasi1999a}.
A Simon-like first mover advantage does not arise
for the Barab{\'a}si-Albert model with a very small seed network
as a new node appears in each time step.

In sum, we have shown through simulations and analysis 
that Simon's fundamental rich-get-richer model---now 60 years old---carries
an intrinsic first-mover advantage.
The first group's size in the idealized model
is outsized by a factor of $1/\innovationrate$,
potentially several orders of magnitude.
Any attempt to attribute Simon's mechanism to
the growth of real-world systems must take into
account this potentially dominant feature of the model,
along with the complications of the variability 
of final rankings for later arriving
groups~\cite{krapivsky2001a,newman2009b,newman2014a,berset2013a,fotouhi2013a},
and past work must come under a new scrutiny.

\smallskip
\acknowledgments
Our paper has benefited from a number of interactions.
We are grateful for discussions with 
Joshua S.\ Weitz,
Jake M.\ Hofman, 
Kameron Decker Harris,
Ryan J.\ Gallagher,
John H.\ Ring IV,
Brian F.\ Tivnan,
and 
Steven H.\ Strogatz.
We also greatly appreciate
collegial correspondence 
in response to our arXiv submission from
Sidney Redner,
James P.\ Gleeson,
Mason A.\ Porter,
Cesar A.\ Hidalgo,
Matus Medo,
and
Mark E.\ J.\ Newman.
Our paper emerged from the ongoing
development of 
PSD's Principles of Complex Systems course
at the University of Vermont.

\clearpage

\end{document}